\documentstyle[12pt]{article}
\begin{document}
\title{Motion of a spin $1\over 2$ particle in shape invariant
scalar and magnetic fields}
  \author{V. M. Tkachuk and P. Roy$^{\dag}$\\
  {\small  Ivan Franko Lviv State University, 
         Department of Theoretical Physics }\\
        {\small 12 Drahomanov Str., Lviv UA--290005, Ukraine}\\
        {\small E-mail: tkachuk@ktf.franko.lviv.ua}\\
 {\small $^{\dag}$  Physics and Applied Mathematics
 Unit,}\\
        {\small Indian Statistical Institute, Calcutta 700035, India}\\
        {\small E-mail: pinaki@isical.ac.in}}

\maketitle

\begin{abstract}
We study the motion of a spin $1\over 2$ particle in a scalar as well as a
magnetic field within the framework of supersymmetric 
quantum mechanics(SUSYQM).
We also introduce the concept of shape invariant scalar and magnetic fields
and it is shown that the problem admits exact analytical solutions when such
fields are considered.


\end{abstract}

\section{Introduction}
The concept of supersymmetry in quantum mechanical models was
first introduced by Nicolai \cite{Nic}. A few years later Witten introduced
supersymmetric quantum mechanics \cite{Wit} as a laboratory 
to examine supersymmetry
breaking in quantum field theoretical models. Subsequently SUSYQM has proved
to be interesting on its own and has been studied by many authors 
from different
points of view \cite{Coop,Jun}.

Over the years it has been shown \cite{Coop,Jun} that SUSYQM plays 
an important
role in obtaining exact solutions of quantum mechanical problems. In fact all
solvable problems of quantum mechanics are either supersymmetric or can be
made so. Now among the various exactly solvable potentials there is a certain
class of potentials which are characterized by a property known as shape
invariance \cite{Gen}.
Potentials
which are shape invariant satisfy certain conditions and it has been shown
\cite{Coop,Jun,Gen} that solutions of the Schr\"odinger equation with any
shape invariant potential can
be obtained in a trivial manner without solving the differential equation.
In fact shape invariance is a sufficient condition for exact solvability.

In the present paper our aim is to use the formalism of SUSYQM to study one
dimensional motion of a spin $1\over 2$
particle in the presence of a scalar potential as well as a magnetic field.
It may be noted that SUSYQM has previously been used to study the motion of
a particle in a magnetic field \cite{de,Ho,Hay}. However,
in the present paper the
problem is similar in nature to a coupled channel problem
\cite{Ama,Hau,Andri,Lev,TkaRoy}.
To solve this problem we shall introduce a definition of shape invariance
which will require not only the scalar potential but also the magnetic field
to satisfy certain conditions. Using this shape invariance property we shall
then obtain exact solutions of the
problem of a spin $1\over 2$ particle moving in a scalar potential and a
magnetic field. The organisation of the paper is as
follows: in section 2 we describe the construction of the hamiltonian describing
the motion of a spin $1\over 2$ particle in a scalar potential and a magnetic
field; in section 3 we introduce the shape invariance conditions and use them
to obtain algebraically exact solutions; finally section 4
is devoted to a conclusion.

\section{Supersymmetric approach to the motion of a spin $1\over 2$ particle
on the real line}

In Witten's model of SUSY quantum mechanics the Hamiltonian
consists of two factorized Schr\"{o}dinger operators

\begin{equation} \label{H}
H_\mp(z;\gamma) = A^\pm(z;\gamma) A^\mp(z;\gamma) = 
-\frac{d^2}{dz^2} + W^2(z;\gamma) \mp W'(z;\gamma)
\end{equation}
where $\gamma$ denotes a set of parameters and the operators 
$A^+(z;\gamma)$ and
$A^-(z;\gamma)$ are given by

\begin{equation}
A^{\pm}(z;\gamma) = \mp\frac{d}{dz} + W(z;\gamma),
\end{equation}
where the function $W(z;\gamma)$ is called the superpotential.

The pair of Hamiltonians in (\ref{H}) are called SUSY 
partner Hamiltonians and
each of these Hamiltonians describe the motion of a spinless particle in
one dimensional potentials 
$V_{\pm}(z;\gamma) = W^2(z;\gamma){\pm} W'(z;\gamma)$. 
Among the
various potentials $V_\pm(z;\gamma)$ those which satisfy the relation
\begin{equation} \label{Shape}
V_+(z;\gamma) = V_-(z;\gamma_1) + \epsilon_1,
\end{equation}
where $\gamma_1=f(\gamma)$ is a function of $\gamma$ and 
$\epsilon_1$ is a constant, are
called shape invariant potentials \cite{Gen}. The shape invariant potentials
are always exactly solvable and their solutions can be obtained purely
algebraically.

We shall now generalize Witten's model of SUSY quantum mechanics in such
way
that each of the Hamiltonians $H_-$, $H_+$ will describe the motion
of a spin
$1\over 2$
particle in a magnetic field and a scalar potential.
In order to do this we generalise the operators $A^{\pm}$ in the following
way:
\begin{equation} \label{Agen}
A^{\pm}(z;\gamma,\beta)=\mp{d\over dz}+W(z;\gamma)+{\bf V}(z;\beta){\bf S},
\end{equation}
It may be noted that here we consider motion of the particle along
$z$-axis and
components of the spin operator ${\bf S}$ are
$S_{\alpha}=\sigma_{\alpha}/2$ ($\alpha=x, y, z$),
$\sigma_{\alpha}$ being the Pauli matrices. Then SUSY partner
Hamiltonians can be
obtained as in (1) and are given by
\begin{equation}\label{Hpm}
H_{\pm}(z;\gamma,\beta)=
-{d^2\over{dz^2}}+V_{\pm}(z;\gamma,\beta)+
{\bf B_{\pm}}(z;\gamma,\beta){\bf S},
\end{equation}
where
\begin{eqnarray}
V_\pm (z;\gamma,\beta)=W^2(z;\gamma)\pm W'(z;\gamma)+ V^2(z;\beta)/4,\\
{\bf B}_\pm (z;\gamma,\beta)=
2W(z;\gamma){\bf V}(z;\beta)\pm{\bf V'}(z;\beta).
\end{eqnarray}
The Hamiltonians $H_\pm$ in (\ref{Hpm}) describe a spin $1\over 2$
particle moving along the
$z$-axis in a scalar potential $V_\pm(z;\gamma,\beta)$ and a magnetic field
${\bf B}_\pm (z;\gamma,\beta)$.

The SUSY Hamiltonian reads
\begin{equation} \label{hamil}
H=
\left(
\begin{array}{c}
H_+ \ \  0 \\
0 \ \ \ H_-
\end{array}
\right)
=\{Q^+,Q_-\},
\end{equation}
where the superchages $Q^+$ and $Q_-$ have the form
\begin{equation}
Q^+= A^-\otimes \sigma^+, \ \
Q^-= A^+\otimes \sigma^-.
\end{equation}

The supercharges and SUSY Hamiltonian fulfil well known $N=2$ SUSY
algebra
\begin{equation}
\{Q^+,Q^-\}=H, \ \ [Q^\pm, H]=0, \ \ (Q^\pm)^2=0.
\end{equation}
Note that in the present case the SUSY Hamiltonian and supercharges
act on four component wave function.
The standard Witten model of SUSY quantum
mechanics can be reproduced by setting $ {\bf V}=0$.

The Hamiltonians $H_+$ and $H_-$ have exactly the same
energy levels (perhaps with the exception of the zero energy state).
For the zero energy ground state the following scenarios are possible:
(1) Zero energy ground state does not exist (broken SUSY)
(2) Zero energy ground state exists for one of the Hamiltonians
$H_-$ or $H_+$ (exact SUSY)
(3) Zero energy ground state exists for both Hamiltonians
$H_-$ and $H_+$ (exact SUSY).
In a previous paper \cite{TkaRoy} it was shown that this last scenario
can be realised when a particle moves in a rotating magnetic field
and a zero scalar potential. For the standard Witten model of SUSY
quantum mechanics such a situation is arises when the superpotential is
a periodic function \cite{DunF,DunM}.

In present paper we shall consider the case when zero energy ground state
exists for one of the Hamiltonians $H_-$ or $H_+$, say for $H_-$.
In this case the eigenvalues $E_n^\pm$ and eigenfunctions
$\psi_n^\pm$ of the Hamiltonians $H_\pm$ are related by the
following SUSY transformations:
\begin{eqnarray} \label{trE}
E^-_{n+1}=E^+_n, \ \ E^-_0=0, \\ \label{tr1psi}
\psi^-_{n+1}={1\over \sqrt{E^+_n}}A^+\psi^+_n, \\
\psi^+_{n}={1\over \sqrt{E^-_{n+1}}}A^-\psi^-_{n+1}. \label{tr2psi}
\end{eqnarray}

In equations (\ref{tr1psi}) and (\ref{tr2psi}) the operators $A^\pm$ are (2 x 2) matrices and the wave
functions $\psi_n^\pm$ are 2 component wave functions. As a result equations
(\ref{tr2psi}) and (\ref{tr2psi}) are matrix differential equations although in appearance they look similar
to the standard SUSY transformations \cite{Coop,Jun}.

\section{Shape invariant potentials and magnetic fields}
In this section we shall generalise the idea of shape invariance
for obtaining exact solution of the eigenvalue problem for a spin $1\over 2$ particle
moving in both a scalar potential as well as a magnetic field.
We begin with the eigenvalue problem corresponding to the Hamiltonian $H_-$
with superpotentials $W=W(z,\gamma)$, ${\bf V}(z,\beta)$
which depend on some parameters $\gamma$ and $\beta$.
Since the zero energy ground state of this Hamiltonian is annihilated by the
operator $A^-(z;\gamma,\beta)$ we have
\begin{equation} \label{psi0zgb}
A^-(z;\gamma,\beta)\psi^-_0(z;\gamma,\beta)=0.
\end{equation}
Note that in the case of standard Witten model of SUSY quantum
mechanics this equation is a single first order differential equation and can be easily
solved.
But in the present case the operator $A^-$ is a (2 x 2) matrix differential operator
and therefore the above equation is a set of two first order coupled differential
equations.
So in general the ground state can not be obtained in terms of the
superpotentials. It is similar to the situation in SUSY quaternionic
quantum mechanics \cite{Dav}.
We shall return to the problem of determining the ground state
later. For the time being let us assume that we have a solution of equation
(\ref{psi0zgb}).

Now let us consider the SUSY partner of $H_-(z;\gamma,\beta)$, i.e. $H_+(z;\gamma,\beta)$.
If we calculate the ground state of $H_+(z;\gamma,\beta)$ we immediately find the first
excited state of $H_-(z;\gamma,\beta)$ using the SUSY transformations
(\ref{trE}) - (\ref{tr2psi}).
Now in order to calculate the ground state of $H_+$ let us rewrite
it in the form
\begin{equation} \label{spH}
H_+(z;\gamma,\beta)=H_-(z;\gamma_1,\beta_1)+\epsilon_1=
A^+(z;\gamma_1,\beta_1)A^-(z;\gamma_1,\beta_1)+\epsilon_1, \ \ \epsilon_1>0,
\end{equation}
where $\epsilon_1$ is the factorisation energy.
The operators $A_\pm(z;\gamma_1,\beta_1)$
corresponding to $H_-(z;\gamma_1,\beta_1)$
have the same form as in (\ref{Agen}) but with superpotentials
$W=W(z,\gamma_1)$, ${\bf V}={\bf V}(z,\beta_1)$.

We note that the wave function of the ground state of $H_+(z;\gamma,\beta)$ is also the wave function of 
ground
state of $H_-(z;\gamma_1,\beta_1)$
i.e. $\psi^+_0(z,\gamma, \beta)=\psi^-_0(z,\gamma_1,\beta_1)$
and it satisfies the equation
\begin{equation}
A^-(z;\gamma_1,\beta_1) \psi^-_0(z,\gamma_1,\beta_1)=0.
\end{equation}

Using the SUSY transformations we can now obtain the energy level
and corresponding wave function of the first excited state of the
Hamiltonian $H_-(z;\gamma,\beta)$:
\begin{equation}
E^-_1=\epsilon_1, \ \ 
\psi^-_1={1\over\sqrt{\epsilon_1}}A^+(z,\gamma,\beta)
\psi^-_0(z,\gamma_1,\beta_1).
\end{equation}

From (\ref{spH}) we can now obtain the conditions of shape invariance
involving the superpotential and the magnetic field. In explicit form
these conditions read
\begin{equation}
\begin{array}{lcl}
W^2(z;\gamma)+W'(z;\gamma)+{\bf V}^2(z;\beta)/4
&=&W^2(z;\gamma_1)-W'(z;\gamma_1)\\
&&+{\bf V}^2(z;\beta_1)/4+\epsilon_1, \label{Scal}\\
\end{array}
\end{equation}
\begin{equation}\label{Mag}
2W(z;\gamma){\bf V}(z;\beta)+{\bf V}'(z;\beta)=
2W(z;\gamma_1){\bf V}(z;\beta_1)-{\bf V}_1'(z;\beta_1).
\end{equation}
Equation (\ref{Scal}) is the condition for shape invariant scalar
superpotential while (\ref{Mag}) is the equation for shape invariant magnetic
field. Comparing with equation (\ref{Shape}) we find that in the present case
shape invariance conditions consist of four equations rather 
than a single one.

In general it is very difficult to solve these equations for superpotentials
(magnetic fields) when an arbitrary magnetic field
(superpotential) is prescribed.
However when we consider some specific superpotential and
magnetic field solutions of equations (\ref{Scal}) and (\ref{Mag}) can still
be obtained.
To this end let us choose {\bf V} in the form
\begin{equation}\label{V}
{\bf V}=g(z){\bf a}+\beta{\bf b},
\end{equation}
where $\bf a$ and $\bf b$ are perpendicular unit vectors i.e,
${\bf ab}=0$.
Then equation (\ref{Scal}) reads
\begin{equation} \label{Scal1}
W^2(z;\gamma)+W'(z;\gamma)+{\beta}^2/4=W^2(z;\gamma_1)-W'(z;\gamma_1)+
{\beta}_1^2/4+\epsilon_1. \label{Scaln}
\end{equation}
and the vector equation (\ref{Mag}) splits into two scalar equations
\begin{eqnarray}
2W(z;\gamma)g(z)+g'(z)=2W(z;\gamma_1)g(z)-g'(z), \label{Mag1}\\
W(z;\gamma)\beta =W(z;\gamma_1)\beta_1. \label{Mag2}
\end{eqnarray}
Then from (\ref{Mag1}) we obtain
\begin{equation} \label{g}
g(z)=\lambda e^{\int^z (W(z;\gamma_1)-W(z;\gamma))},
\end{equation}
where $\lambda$ is some constant.
Here it is important to note that since $g(z)$ does not depend on the
parameters $\gamma$ the difference between the new and the old superpotential
$(W(z;\gamma_1)-W(z;\gamma))$ also does not depend on these parameters.

In order to satisfy equation (\ref{Mag2}) we now choose
\begin{equation} \label{W}
W=\gamma f(z),
\end{equation}
which leads to the following relation between the parameters
\begin{equation} \label{gambet}
\gamma\beta=\gamma_1\beta_1.
\end{equation}

Note that only superpotentials of the form (\ref{W}) ensures that the
difference $(W(z;\gamma_1)-W(z;\gamma))$ is independent of the parameter $\gamma$.
Thus the superpotentials (\ref{V})
and (\ref{W}) lead to shape invariant scalar potential and
magnetic field and so the corresponding eigenvalue problem can be solved
exactly.

To find the exact solutions we now continue the shape invariant construction
recursively and obtain the
energy levels and the corresponding wave functions
of $H_-$ in the following form
\begin{eqnarray} \label{En}
E^-_n =\sum^{n}_{i=0}\epsilon_i,
\label{En}\ \ \epsilon_0=0,  \\ \nonumber
\psi^-_n(z;\gamma,\beta)= \\ \label{psin}
C^-_n A^+(z;\gamma,\beta)...
A^+(z;\gamma_{n-2},\beta_{n-2})A^+(z;\gamma_{n-1},\beta_{n-1})
\psi^-_0(z; \gamma_n, \beta_n),
\end{eqnarray}
where
$C^-_n$ are normalization constants,
$\psi^-_0(z;\gamma_n,\beta_n)$ is the zero energy eigenfunction of
$H_-(z;\gamma_n,\beta_n)$ which satisfies the equation
$A^-(z;\gamma_n,\beta_n)\psi^-_0(z;\gamma_n,\beta_n)=0$,
$A^\pm(z;\gamma_n,\beta_n)$
and $H_-(z;\gamma_n,\beta_n)$ are of the form (\ref{Agen}) and (\ref{Hpm})
respectively
with superpotentials $W(z;\gamma_n)$, ${\bf V}(z;\beta_n)$.
In our notations
$\gamma_0=\gamma$ and $\beta_0=\beta$.

In explicit form the equation determining $\psi^-_0(z,\gamma_n,\beta_n)$ reads
\begin{equation} \label{psi0n}
\left({d\over dz} + \gamma_n f(z) + (g(z){\bf a}+\beta_n{\bf b})
{\bf S}\right)\psi^-_0(z, \gamma_n, \beta_n) =0.
\end{equation}
The superpotential $\gamma_n f(z)$ can be eliminated from this equation by
using the following
transformation
\begin{equation} \label{psi}
\psi^-_0(z, \gamma_n, \beta_n) = \phi(z,\beta_n)
e^{-\int \gamma_n f(z)},
\end{equation}
where $\phi$ is a two component function which satisfies the equation
\begin{equation} \label{Eqphi}
\left({d\over dz} + (g(z){\bf a}+\beta_n{\bf b})
{\bf S}\right)\phi(z, \beta_n) =0.
\end{equation}

Let us now choose $\bf a$ parallel to z-axis, $\bf b$ parallel to x-axis.
Then equation (\ref{Eqphi}) which is a set of two first order coupled differential equations can
be rewritten in the form
\begin{eqnarray}
a^-\phi_1(z,\beta_n)=-{\beta_n\over 2} \phi_2(z, \beta_n), \label{am} \\
a^+\phi_2(z, \beta_n)={\beta_n\over 2}\phi_1(z, \beta_n), \label{ap}
\end{eqnarray}
where the operators $a^\pm$ are given by
\begin{equation}
a^\pm=\mp{d\over dz}+g(z)/2.
\end{equation}
The above set of first order coupled equations can easily be transformed
into second order equations for $\phi_1$ and $\phi_2$ and are given by
\begin{eqnarray}
a^+a^-\phi_1(z, \beta_n) = h_-\phi_1
=-{\beta_n^2\over 4}\phi_1(z, \beta_n), \label{apm1} \\
a^-a^+\phi_2(z, \beta_n) = h_+\phi_2
=-{\beta_n^2\over 4}\phi_2(z, \beta_n). \label{amp2}
\end{eqnarray}

It is interesting to note that equations (\ref{apm1})
and (\ref{amp2}) have the form of eigenvalue equations corresponding to
$H_\pm$ of one dimensional SUSY quantum mechanics (see equation (\ref{H}))
with superpotential $g(z)/2$ and $-\beta^2_n/4$ can be treated as energy
which is negative in the present case. The solutions of equations
(\ref{apm1}) and (\ref{amp2}) need not necessarily be square integrable functions.
However nonsquare integrable solutions of (\ref{apm1}) and (\ref{amp2}) can
still be used to obtain physical solutions of the original eigenvalue
equation (see equation (\ref{psi})).

\subsection{Examples}

{\bf Case 1}: The simplest superpotential which we can choose is $W=\gamma z$.
But in this case 
from (\ref{Scal1}) it follows that $\gamma_1=\gamma$ and
we find from (\ref{g}) that $g=const$. As a result we obtain
a magnetic field which does not change its direction.
Therefore this case can be reduced to the standard Witten model of
SUSY quantum mechanics.

{\bf Case 2}: Let us now consider the following superpotential:

\begin{equation}
W=\gamma \tanh (z), \ \ \gamma >0.
\end{equation}
Then iterating the shape invariant condition (\ref{Scal1}) $n$ times we obtain
\begin{eqnarray}
\epsilon_n=\gamma^2_{n-1}-\gamma^2_n+(\beta^2_{n-1}-\beta^2_n)/4,
\label{Epsn}\\
\gamma_{n-1}(\gamma_{n-1}-1)= \gamma_n(\gamma_n+1). \label{Eqngamma}
\end{eqnarray}
Equation (\ref{Eqngamma}) have two solutions with respect to $\gamma_n$.
But only one of them is acceptable from the point of view of square
integrability of the wave function and this is given by
\begin{equation}
\gamma_n=\gamma_{n-1}-1=\gamma-n.
\end{equation}
Now iterating the relation (\ref{gambet}) $n$ times we obtain
\begin{equation}
\beta_n={\gamma_{n-1}\over \gamma_n}\beta_{n-1}=
{\gamma\over \gamma_n}\beta
\end{equation}

To determine $g(z)$ we use (\ref{g}) and obtain
\begin{equation}
g(z)={\lambda\over \cosh(z)}. \label{gz}
\end{equation}
From (\ref{gz}) it is seen that the function $g(z)$ indeed does not depend on
the parameters appearing in the superpotential.
Now  using $W(z;\gamma)$ and $g(z)$ we can calculate the scalar potential
and the magnetic field in which the spin $1\over 2$ particle is
moving:
\begin{eqnarray}
V_{\pm}={\lambda^2/4-\gamma (\gamma\mp
1)\over\cosh^2(z)}+\gamma^2+\beta^2, \\
{\bf B}_\pm={\lambda^2\over 2}(2\gamma\mp 1){\tanh(z)\over \cosh(z)}{\bf
a}+
2\gamma\beta \tanh(z){\bf b}.
\end{eqnarray}

Now let us study the eigenvalue equations (\ref{apm1}) and (\ref{amp2}).
To determine whether the spectrum is finite or infinite it is now necessary
to establish the maximum value of $n$.
In the present case equation (\ref{apm1}) reads
\begin{equation} \label{Exphi0}
\left( -{d\over dz}+{\lambda\over 2\cosh(z)}\right)
\left( {d\over dz}+{\lambda\over 2\cosh(z)}\right)\phi_1(z, \beta_n)=
-{\beta_n^2\over 4}\phi_1(z, \beta_n).
\end{equation}

The asymptotic behaviour of the solutions of equation (\ref{Exphi0})
at $|z|\to \infty$ is given by
\begin{equation}
\phi_1(z, \beta_n)= {\rm const} e^{\pm\beta_n z/2}. \label{phi1}
\end{equation}
Using (\ref{am}) for second component we obtain
\begin{equation}
\phi_2(z, \beta_n)= \mp{\rm const} e^{\pm\beta_n z/2}. \label{phi2}
\end{equation}
From (\ref{phi1}) and (\ref{phi2}) it is seen that the solutions are not
square integrable.
Now to determine the asymptotic behaviour of $\psi_0 (z,\gamma_n, \beta_n)$
we use (\ref{phi1}) and (\ref{phi2}) in (\ref{psi}) and obtain

\begin{equation} \label{Apsi0n}
\psi^-_0(z, \gamma_n, \beta_n)=
{\rm const}
\left(
\begin{array}{c}
1\\\mp 1
\end{array}
\right)
{e^{\pm\beta_n z/2}\over \cosh^{\gamma_n}(z)}.
\end{equation}
Then from the condition of square integrability of $\psi^-_0(z, \gamma_n, \beta_n)$
we get
\begin{equation}
\gamma_n >|\beta_n|/2. \label{cond}
\end{equation}
Thus it follows from (\ref{cond}) that
\begin{equation}
n<\gamma -\sqrt{\gamma |\beta| /2}.
\end{equation}

Energy levels of the Hamiltonian $H_-(z;\gamma,\beta)$ are then given by
\begin{equation} \label{ExEn}
E_n^-=\gamma^2 -(\gamma -n)^2 +
{\beta^2\over 4}\left( 1-{\gamma^2\over(\gamma -n)^2}\right).
\end{equation}

From (\ref{Apsi0n}) it follows that there are two independent square integrable solutions
of equation (\ref{psi0n}) and as a result the ground state of $H_-(z;\beta_n,\gamma_n)$ is twofold
degenerate. Now from (\ref{psin}) and (\ref{psi0n}) it can be shown that each
energy level $E^-_n$ is doubly degenerate.

We now proceed to determine the eigenfunctions of $H_-(z;\beta_n,\gamma_n)$ in explicit form.
In order to do this we need to have the general solutions of equation (\ref{Exphi0}).
To solve this equation we first transform it to the equation for hypergeometric
functions.
Let us introduce a new variable $x=\sinh (z)$. Then equation (\ref{Exphi0})
becomes
\begin{equation}
\left[
-(1+x^2){d^2\over dx^2}-x{d\over dx}+{\lambda^2\over 4}{1\over 1+x^2}
+{\lambda\over 2}{x\over 1+x^2}
\right]\phi_1=-{\beta_n^2\over4}\phi_1.
\end{equation}

We now introduce a new function $f$ defined by the relation
\begin{equation}
\phi_1=fe^{-{\lambda \over 2}\arctan(x)}
\end{equation}
and use a new variable $\xi=(1-ix)/2$ to obtain from equation (\ref{Exphi0})

\begin{equation}
\left[
(1-\xi)\xi{d^2\over d\xi^2}+\left({1\over 2}-i{\lambda\over 2}
-\xi\right){d\over d\xi}
\right]f=-{\beta_n^2\over4}f.
\end{equation}

This equation has two linearly independent solutions:
\begin{eqnarray}
f^{(1)}=F(a,b;c;\xi), \\
f^{(2)}=\xi^{1-c}(1-\xi)^{c-a-b}F(1-a,1-b;2-c;\xi),
\end{eqnarray}
where F(a ,b ;c ;x) is the hypergeometric function and
$a=\beta_n/2$, $b=-\beta_n/2$, $c=1/2-i\lambda /2$.

Then
using (\ref{psin}) we obtain in explicit form
two eigenfunctions which correspond to
the same energy level given by (\ref{ExEn}).
As a result we conclude once more
that energy levels of  $H_-(z;\beta,\gamma)$ are twofold degenerate.

Now let us analyze the reason for this double degeneracy of the energy levels
of $H_-(z;\gamma,\beta)$.
Note however that this double degeneracy is not related to the
SUSY of original Hamiltonian which consists of two
partner Hamiltonians $H_-(z;\gamma,\beta)$ and $H_+(z;\gamma,\beta)$.
The degeneracy of $H_-(z;\gamma,\beta)$ is related to the spin degrees
of freedom of the Hamiltonian and also with the existence of an additional
integral of motion $T=I\sigma_y$ in the case when $W(-z)=-W(z)$,
where $I$ is parity operator and acts according to
$If(z)=f(-z)$.
Also $T^2=1$ and thus this operator has two eigenvalue $\pm1$.
We also note that the operator of complex conjugation $R$,
acting according to $Rf=f^*$,
commutes with
$A^\pm(z;\gamma_i,\beta_n)$ in the case when
$\bf a$ is parallel z axis and $\bf b$
is parallel x axis.
These operators satisfy the (anti)commutation relations
\begin{eqnarray} \label{antTA}
TA^\pm(z;\beta_n,\gamma_n) + A^\pm(z;\beta_n,\gamma_n) T=0, \\ \label{antTR}
TR+RT=0, \\ \label{antRA}
R A^\pm(z;\beta_n,\gamma_n) - A^\pm(z;\gamma_n,\beta_n) R =0
\end{eqnarray}
Furthermore the operators $R$ and $T$ commute with the Hamiltonian
$H_-(z;\gamma_n,\beta_n)$
\begin{equation}
[T,H_-(z;\gamma_n,\beta_n)]=[R,H_-(z;\gamma_n,\beta_n)]=0. \label{commute}
\end{equation}

Let us now demonstrate using the above algebra 
that zero energy level for
$H_-(z;\gamma_n,\beta_n)$ is doubly degenerated.
To show this let us suppose that we have at least one zero energy ground
state. As a result of the commutation relation (\ref{commute})
this state can be chosen also as eigenfunction of the operator $T$.
Thus the zero energy ground state satisfies the equations
\begin{eqnarray}
T\psi_\lambda =\lambda\psi_\lambda, \label{EqT}\\
A^-\psi_\lambda =0, \label{EqAzero}
\end{eqnarray}
where $\lambda$ take one of the values $1$ or $-1$.
Then from (\ref{antTR}) and (\ref{EqT})
it follows that $R\psi_\lambda=\psi_{-\lambda}$
Now operating $R$ from the left on (\ref{EqAzero}) and using (\ref{antRA})
we obtain
\begin{equation}
A^-R\psi_\lambda = A^-\psi_{-\lambda}=0.
\end{equation}
Thus $\psi_{-\lambda}$ together with  $\psi_{\lambda}$
are wave functions of the zero energy
ground state. We can conclude that zero energy level
of the Hamiltonian $H_-(z;\gamma_i,\beta_i)$ is doubly degenerate.
Since the $n$th excited state of the Hamiltonian $H_-(z;\gamma,\beta)$
is related by (\ref{psin}) to
the ground state of the Hamiltonian $H_-(z;\beta_n,\gamma_n)$ 
we conclude that all the energy levels 
of the Hamiltonian $H_-(z;\gamma,\beta)$
are doubly degenerate. Finally we note that as a consequence of the relation
(\ref{trE}) the zero energy ground level of the full Hamiltonian (\ref{hamil}) 
is doubly
degenerate while the excited levels are fourfold degenerate. 

\section{Conclusions}
In the present paper we extend the definition of shape invariance to obtain
exact solutions of the eigenvalue
problem relating to the motion of a spin $1\over 2$ particle moving in scalar
potential and a
magnetic field. The shape invariance conditions are more
complicated than in the standard case. 
It is because 
instead of one
equation for the superpotential $W$ in standard case
we have four equations coupling the
superpotential with the components of the vector function ${\bf V}$.
It has been shown that if we choose a superpotential and a magnetic field
satisfying the above mentioned shape invariance condition we can obtain
exact analytical solutions of the eigenvalue problem. The spectrum of the
full Hamiltonian is fourfold degenerate while those of the component
Hamiltonians are doubly degenerate. We have also analysed the reasons of
this double degeneracy and it has been shown to be due existence of
additional integrals of motion rather than SUSY. We feel it would be of
interest to find other
superpotentials and magnetic fields which are shape invariant and are thus
exactly soluble.

\end{document}